\documentclass[
showkeys,
reprint,
superscriptaddress,
nofootinbib,
nobibnotes,
longbibliography,
aps,
prb,
floatfix,
nofootinbib
]{revtex4-1}

\usepackage{mathptmx}
\usepackage{amsmath}
\usepackage{amssymb}
\usepackage{graphicx}
\usepackage{dsfont}
\usepackage[dvipsnames]{xcolor}
\usepackage[breaklinks=true,colorlinks=true,linkcolor=blue,urlcolor=blue,citecolor=blue]{hyperref}

\begin{document}

\title{Microwave generation and vortex jets in superconductor nanotubes}

\author{I. \surname{Bogush}}
\affiliation{Institute for Emerging Electronic Technologies, Leibniz IFW Dresden, Helmholtzstraße 20, 01069 Dresden, Germany}
\affiliation{Moldova State University, Str. A. Mateevici 60, 2009 Chi\c{s}in\u{a}u, Republic of Moldova}
\author{O. V. \surname{Dobrovolskiy}}
\affiliation{University of Vienna, Faculty of Physics, Nanomagnetism and Magnonics, Superconductivity and Spintronics Laboratory, W\"ahringer Str. 17, 1090 Vienna, Austria}
\author{V. M. \surname{Fomin}}
\email{v.fomin@ifw-dresden.de}
\affiliation{Institute for Emerging Electronic Technologies, Leibniz IFW Dresden, Helmholtzstraße 20, 01069 Dresden, Germany}
\affiliation{Moldova State University, Str. A. Mateevici 60, 2009 Chi\c{s}in\u{a}u, Republic of Moldova}

\begin{abstract}
The dynamics of magnetic flux quanta (Abrikosov vortices) determine the resistive response of superconductors. In pinning-free planar thin films, the penetration and motion of vortices are controlled by edge defects, leading to such arrangements as vortex chains, vortex jets, and phase-slip regimes. Here, relying upon the time-dependent Ginzburg-Landau equation, we predict that these vortex patterns should appear in superconductor open nanotubes even without edge defects, due to the inhomogeneity of the normal magnetic induction component $B_\mathrm{n}$, caused by the 3D tube geometry. The crossing of the half-tubes by dc-driven vortices induces GHz-frequency voltage $U$ oscillations with spectra $U_\mathrm{f}(B)$ evolving between $nf_1$ and $\frac{n}{m}f_1$ [$f_1$: vortex nucleation frequency; $n,m\geq 2$] and blurred in certain ranges of currents and fields. An $nf_1$-spectrum corresponds to a single vortex-chain regime typical for low $B$ and for tubes of small radii. At higher fields, an $\frac{n}{m}f_1$-spectrum points to the presence of $m$ vortex chains in the vortex jets which, in contrast to planar thin films, are not diverging because of constraint to the tube areas where $B_\mathrm{n}$ is close to maximum. A blurry spectrum implies complex arrangements of vortices because of multifurcations of their trajectories. Finally, due to a stronger confinement of single vortex chains in tubes of small radii, we reveal peaks in $dU/dB$ and jumps in the frequency of microwave generation, which occur when the number of fluxons moving in the half-tubes increases by one. In all, our findings are essential for novel 3D superconductor devices which can operate in few- and multi-fluxon regimes.
\end{abstract}


\maketitle

\section{Introduction}
Superconductivity and dynamics of magnetic flux quanta (Abrikosov vortices, or fluxons) in open nanotubes possess a number of features distinct from those in planar thin films\,\cite{Fom12nal}. For instance, the curved geometry of a nanotube leads to inhomogeneous normal, $\mathbf{B}_\mathrm{n}$, and tangential, $\mathbf{B}_\mathrm{t}$, components of the magnetic induction\,\cite{Fom22apl}, which decisively affects the edge barriers\,\cite{Zel94prl,Fri01prb,Mik21prb} for vortex nucleation and the topology of screening currents\,\cite{Fom22nsr}. As known, vortices tend to nucleate and consequently reside in the regions with suppressed superconductivity\,\cite{thinkham1996}. In planar structures, edge defects suppress the energy barrier for the nucleation of vortices and act as gates for the vortex nucleation\,\cite{Ala01pcs,Vod03pcs,fomin2009vortex,Cle11prb}. In pinning-free planar thin films, in the presence of a dc transport current of density $\mathbf{j}_\mathrm{tr}$, which exceeds the critical current associated with the edge-barrier suppression, the vortices nucleating at an edge defect are not constrained to cross the film strictly perpendicular to $\mathbf{j}_\mathrm{tr}$. Instead, the vortices tend to arrange themselves into a \emph{vortex jet} which is narrow near the defect and \emph{expands} due to the vortex-vortex repulsion as they move to the opposite edge of the constriction\,\cite{Bez22prb}. By contrast, since in 3D structures superconductivity is depressed in the regions close to maximum $\mathbf{B}_\mathrm{n}$, this symmetry breaking determines the places for vortex nucleation, and -- as will be shown in this work -- vortex jets consisting of a few vortex chains become constrained to move \emph{parallel} to the tube axis.

When a vortex array moves in a superconductor, it induces ac voltage oscillations\,\cite{Kul66spj} while its reaching the superconductor boundary leads to electromagnetic radiation\,\cite{Bul06prl}. Previously, such a GHz-frequency generation was experimentally observed for vortices moving in Mo/Si superlattices and Nb thin films using microwave pickup antennas\,\cite{Dob18apl,Dob18nac}. Furthermore, periodic nucleation of vortices at a sample edge at frequencies of tens of GHz was experimentally demonstrated by scanning SQUID-on-tip microscopy for Pb thin film constrictions\,\cite{Emb17nac}. Correlated dynamics of vortices were also studied via electrical resistance measurements, revealing ac/dc quantum interference effects related to the periodicity of the vortex lattice\,\cite{Fio71prl,Mar75ssc} and kinks in the current-voltage curves when the number of fluxons in the constriction increased by one\,\cite{Asl75etp,Niv93prl,Ust23etp,Bev23pra}.

However, the aforementioned studies were performed so far for planar superconductor structures. At the same time, the maturity of advanced nanofabrication techniques has recently made available 3D superconductor nanoarchitectures featuring curved geometries and complex topologies\,\cite{Fom21boo,Mak22adm}. For instance, direct writing by focused particle beams\,\cite{Fer20mat,Orus2021,Hof23arx} and the self-rolling technology\,\cite{Thurmer08,Thurmer10,Loe19acs} make accessible superconductors shaped as tubes and helices and therefore urge theoretical studies of their magneto-transport properties. While time- and spatially-resolved experimental studies of vortex dynamics in 3D nanoarchitectures are challenging\,\cite{Mak22adm}, deduction of vortex configurations from an analysis of global observables (average induced voltage, its oscillatory component and frequency spectrum) represents one of the viable approaches to that end.

Here, we investigate theoretically the dynamics of vortices in 3D superconductor open nanotubes (nanotubes with a slit) exposed to an azimuthal transport current and an external magnetic field directed perpendicular to the tube axis. We predict distinct voltage spectra $U_\mathrm{f}$ for the various regimes in the vortex dynamics, which evolve between $nf_1$ and $\frac{n}{m}f_1$ [$f_1$: vortex nucleation frequency; $n,m\geq 2$] and are blurred in a certain range of currents and fields. Our theoretical predictions can be examined experimentally for, e.g. open Nb nanotubes fabricated by the self-rolling technology\,\cite{Thurmer08,Thurmer10,Loe19acs}. The results are essential for 3D superconductor devices which can operate in few- and multi-fluxon regimes.

The performed numerical modeling is based on the time-dependent Ginzburg-Landau (TDGL) equation. Evolution of the global observables in conjunction with the spatial distribution of the superconducting order parameter reveals the following correlation: A voltage frequency spectrum of the $nf_1$-type corresponds to a single vortex chain regime typical for low $B$ and for tubes of small radii. At higher fields, a $\frac{n}{m}f_1$ voltage spectrum points to the presence of $m$ vortex chains in the vortex jets which, in contrast to planar thin films, are not diverging because of constraint to the tube areas where $B_\mathrm{n}$ is close to maximum. A blurry voltage spectrum implies complex arrangements of vortices emerging because of multifurcations of their trajectories. Furthermore, due to a stronger confinement of single vortex chains in tubes of small radii, we reveal peaks in $dU/dB$ and jumps in the frequency of microwave generation, which occur when the number of fluxons in the half-tubes increases by one.

The paper is organized as follows. After the problem statement in Sec.\,\ref{sec:problem}, in Sec.\,\ref{sec:spectrum} we discuss the voltage spectra in conjunction with the vortex arrangements. In Sec.\,\ref{sec:amp_osc} we examine the oscillatory behavior of the induced voltage as a function of the magnetic field. In Secs.\,\ref{sec:jets} and\,\ref{sec:generation}, we summarize the major theoretical findings and comment on the frequency range of the generation. In Sec.\,\ref{sec:confinement} we demonstrate that a reduction of the nanotube radius results in promotion of single-vortex-chain regimes and noticeable deformations of vortex paths due to the enhanced interaction between vortices in the both half-tubes. Finally, in Sec.\,\ref{sec:planar} we compare the induced voltage spectra for nanotubes with those for planar structures and conclude our presentation in Sec.\,\ref{sec:applic} with an outline of the applicability of the obtained results.

\section{Model}
\label{sec:problem}
\begin{figure}[t!]
    \centering
    \includegraphics[width=0.8\linewidth]{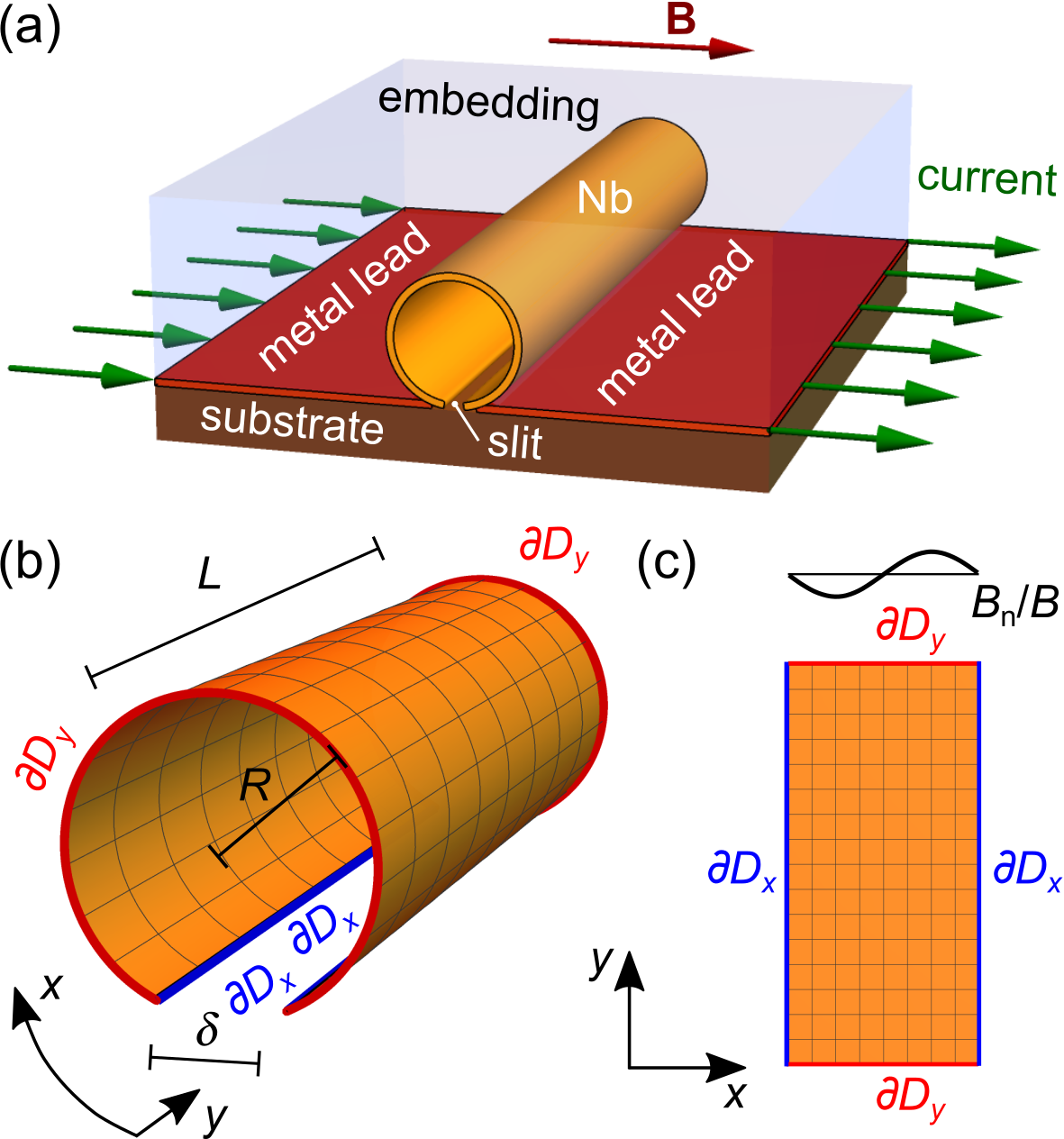}
    \caption{Geometry (a) and the mathematical model (b) of an open nanotube. (c) Unwrapped nanotube surface with the indicated modulation of the normal component of the magnetic induction $B_\mathrm{n}/B$. The normally conducting current (and voltage) leads are attached to the slit banks and correspond to the $\partial D_x$ boundaries.}
    \label{fig:scheme}
\end{figure}

The geometry of the problem is shown in Fig.\,\ref{fig:scheme}(a). A superconductor nanotube of radius $R$, length $L$, with a slit of width $\delta$ is exposed to an azimuthal transport current of density $\mathbf{j}_\mathrm{tr}$. A magnetic field producing the flux density $\mathbf{B}$ is applied perpendicular to the tube axis and parallel to the substrate plane. Under the action of the transport current, vortices move in the paraxial direction within each half-tube. Namely, the vortices nucleate at the free edges of the tube [boundaries $\partial D_y$ in Fig.\,\ref{fig:scheme}(b)], move largely parallel to the tube axis, and denucleate at the opposite free edges. The motion of vortices in the opposite half-tubes occurs in reverse directions, because of the $B_\mathrm{n}$ sign reversal, see Fig.\,\ref{fig:scheme}(c). The task is to calculate the spatially-averaged instantaneous voltage $U(t)$ and its frequency spectrum $U_\mathrm{f}$ as functions of $j_\mathrm{tr}$ and $B$, and to associate the features in these global observables with various arrangements in the moving vortex arrays.

The open-tube geometry requires a note on the terminology. Thus, for planar structures, a transport current can lead to the nucleation of vortices and antivortices at the opposite edges\,\cite{Gla86ltp}. The terms ``vortex'' and ``antivortex'' are based on the direction of the vortex magnetic field so that vortices and antivortices in planar structures possess opposite vorticities and attract each other\,\cite{thinkham1996}. In contradistinction, as the orientation of the applied magnetic field is changed in relation to the nanotube surface, see Fig.\,\ref{fig:scheme}(c), vortices from the both half-tubes exhibit opposite vorticities when mapped to a 2D unwrapped manifold, despite corresponding to the magnetic field of the same original orientation. As a result, vortices from different half-tubes \emph{attract} each other.

The modeling is done relying upon the TDGL equation for two tubes of radii $R = 390$\,nm and $240$\,nm. Both tubes have a length $L=5$\,$\mu$m, a slit width $\delta =60$\,nm and a thickness $d=50$\,nm, such that a current density of $1$\,GA/m$^2$ corresponds to a transport current of $0.25$\,mA. An additional planar structure of width $W=750$\,nm and the same length $L=5$\,$\mu$m is considered further for comparison. Details on the equations, boundary conditions, and materials parameters, which are typical for Nb thin films, are provided in the Appendix. The modeling is performed at temperature $T/T_\mathrm{c} = 0.952$, where $T_\mathrm{c}$ is the superconducting transition temperature.

We denote the period of time between two subsequent vortex nucleations at the same half-tube edge as $\Delta t_\mathrm{n}$ and the time-of-flight for a vortex from an edge to the opposite one as $\Delta t_\mathrm{fl}$. The characteristic times $\Delta t_\mathrm{n}$ and $\Delta t_\mathrm{fl}$ depend on $B$ and $j_\mathrm{tr}$. The motion of vortices induces a voltage between the leads attached to the $\partial D_x$ boundaries. The time-dependent voltage $U(t)$ is calculated as the difference of the electric potential averaged over the entire length of the voltage leads at a certain instant of time $t$. For the calculation of the voltage frequency spectrum $U_\mathrm{f}$ a discrete fast Fourier transform (FFT) is applied. Before the FFT, the initial voltage is multiplied with the Hann function to reduce numerical artifacts\,\cite{nuttall1981some}.

The voltage induced by a moving vortex is proportional to its velocity which, in its turn, is affected by the vortex-edge and vortex-vortex interactions in our model which neglects vortex pinning due to possible structural defects\,\cite{thinkham1996}. Depending on $B$ and $j_\mathrm{tr}$, the patterns of vortex trajectories can look differently, e.g., vortices can traverse the tube along a narrow single path (\emph{vortex chain} regime) or in a staggered manner following many paths (\emph{vortex jet} regime). The regime of vortex motion has an impact on the distances between the vortices and on the induced voltage via their (de)nucleation frequencies. In an effort to retrieve information about the arrangement of vortex paths through the global observables, the experimentally accessible voltage and its frequency spectrum are analyzed next.

\section{Results}
\label{sec:results}

\subsection{Voltage spectrum}
\label{sec:spectrum}

\begin{figure}
    \includegraphics[width=1\linewidth]{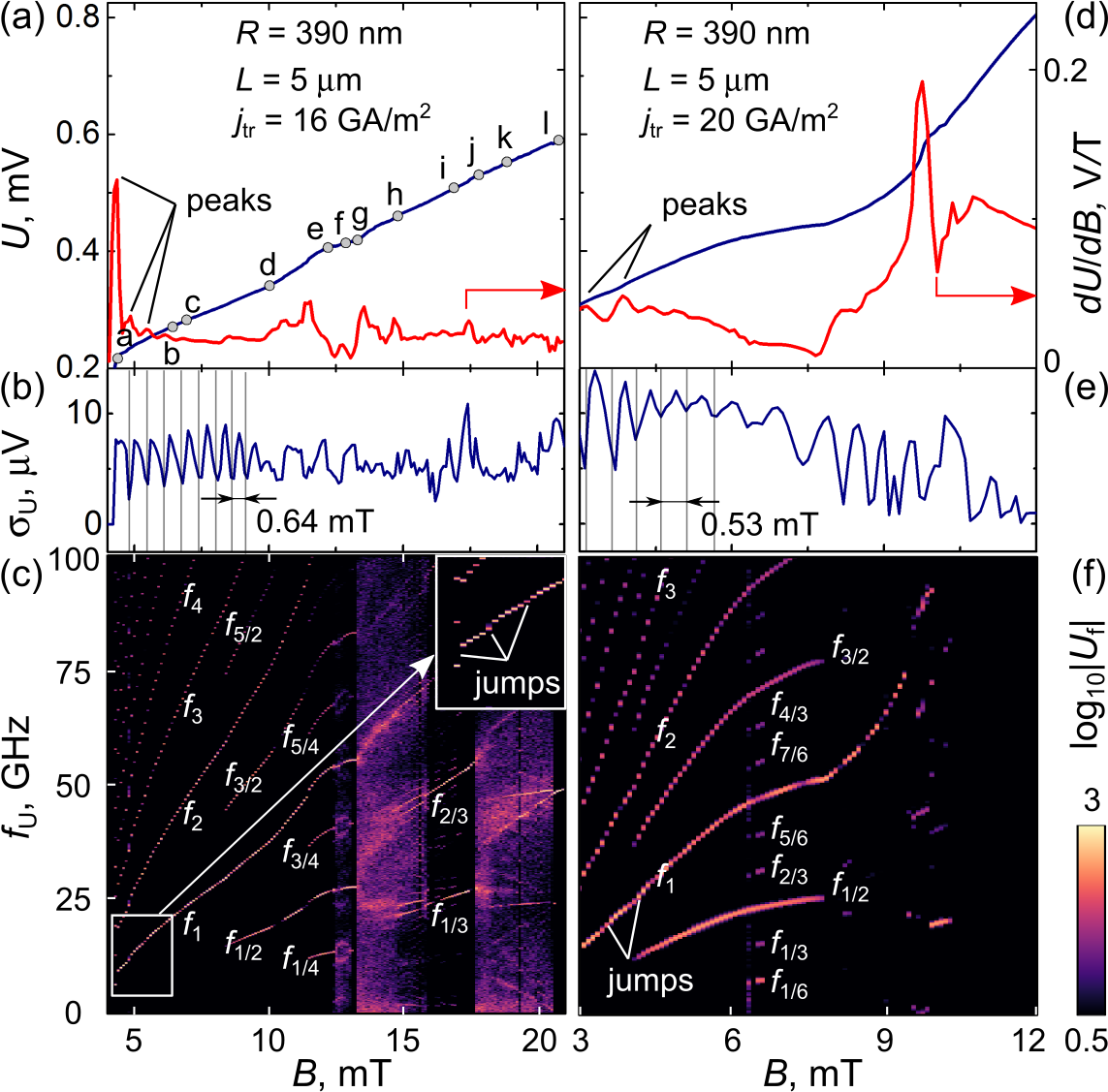}
    \caption{(a,d) Average voltage $U$, (b,e) its root-mean-square deviation $\sigma_U$ and (c,f) frequency spectrum $U_\mathrm{f}$ as functions of the magnetic induction $B$ for the nanotube with $R=390$\,nm at the transport current densities 16\,GA/m$^2$ and 20\,GA/m$^2$.  Points a to l in (a) correspond to the respective panels in Figs.\,\ref{fig:waveforms} and \ref{fig:trajectories}. Solid lines in (b) and (e) illustrate the approximate periodicity of the oscillations of $\sigma_\mathrm{U}(B)$.} \label{fig:spectra}
\end{figure}

Figure\,\ref{fig:spectra} shows the evolution of the average voltage $U$, its root-mean-square deviation (RMSD) $\sigma_\mathrm{U}$ and the frequency spectrum $U_\mathrm{f}$ as functions of $B$ for the nanotube with $R = 390$\,nm at $j_\mathrm{tr}=16$\,GA/m$^2$. While the average voltage increases quasi-linearly with increase of $B$, a few peaks in the derivative $dU/dB$ can be seen in Fig.\,\ref{fig:spectra}(a). The jumps in the frequency $f_\mathrm{U}$ of microwave generation between $4$ and $6.5$\,mT are also enlarged in the inset in Fig.\,\ref{fig:spectra}(c). Prior to discussing Fig.\,\ref{fig:spectra}(b), we outline the main features in the voltage frequency spectrum in Fig.\,\ref{fig:spectra}(c). Three types of frequencies are manifested in $U_\mathrm{f}(B)$: The main frequency $f_1$ is the vortex nucleation frequency, higher harmonics are multiples of the main frequency $f_n = n f_1$ with a positive integer $n$ ($n \ge 2$), and subharmonics are non-integer rational fractions of the main frequency, $f_{n/m} = \frac{n}{m} f_1$ with a positive integer $m$ ($m \ge 2$, $n$ and $m$ have no common divisors). The presence of higher harmonics in $f_\mathrm{U}$ is attributed to the system's nonlinearity, while the occurrence of subharmonics will be explained in what follows.

As follows from Fig.\,\ref{fig:spectra}, vortex motion and microwave generation commence at about $4$\,mT. In the magnetic field range from $4$ to $8$\,mT, the main frequency $f_1$ is accompanied by a series of higher harmonics $f_n$. From $8$ to $12$\,mT, a set of subharmonics $f_q$ appears with $q=1/2,\,3/2,\,\ldots$, followed by another set of subharmonics with $q=1/4,\,3/4,\,\ldots$ between $12$ and $13$\,mT. From $16$ to $17.5$\,mT subharmonics related to $f_{n/3}$ and a very weak subharmonic $f_{1/6}$ are observed. Above $20.5$\,mT, the regime characterized by $f_{1/2}$ re-enters, accompanied by relatively weak occurrences of $f_{1/4}$. Between the aforementioned regimes characterized by well-determined frequencies, the spectrum is blurred without any prominent frequency. Nevertheless, a few frequency ranges can be resolved, forming a fork-like pattern. In the entire range of considered magnetic fields, the average voltage $U$ and the frequency $f_1$ exhibit an almost linear increase with increasing $B$.

Figure\,\ref{fig:spectra}(d-f) shows the evolution of the average voltage, its RMSD and frequency spectrum as functions of $B$ for the same tube at the larger current density $j_\mathrm{tr}=20$\,GA/m$^2$. In this case, nucleation of vortices commences at a smaller field of $2$\,mT, as expected due to a stronger suppression of the edge barrier by the larger transport current. In the voltage frequency spectrum, one recognizes the main frequency, higher harmonics and subharmonics at $f_{n/2}$ and $f_{n/6}$. In contradistinction to the previous case, the spectrum at $j_\mathrm{tr}=20$\,GA/m$^2$
is non-blurry over the entire magnetic field range. We attribute this fact to a stronger-correlated regimes in the vortex dynamics at the larger transport current. However, at $B\geq11$\,mT, the alternating voltage component is weakened significantly due to the formation of a strip with suppressed superconductivity rather then individual vortices in the region near the $B_\mathrm{n}$ maximum. We note that this regime is distinct from the phase slips occurring in the opposite-to-slit region (where $B_\mathrm{n}$ is close to its minimum) unveiled previously for the regimes of fast current or magnetic field sweeps\,\cite{Bog22prb,Rez20cph}. At $B\sim 7$\,mT, the subharmonics $f_{n/6}$  are dominant, while the subharmonics $f_{n/2}$ prevail in the range $4.5$-$8$\,mT. The presence of various sets of harmonics is attributed to different vortex arrangements in the half-tubes.

\begin{figure}
    \centering
    \includegraphics[width=0.86\linewidth]{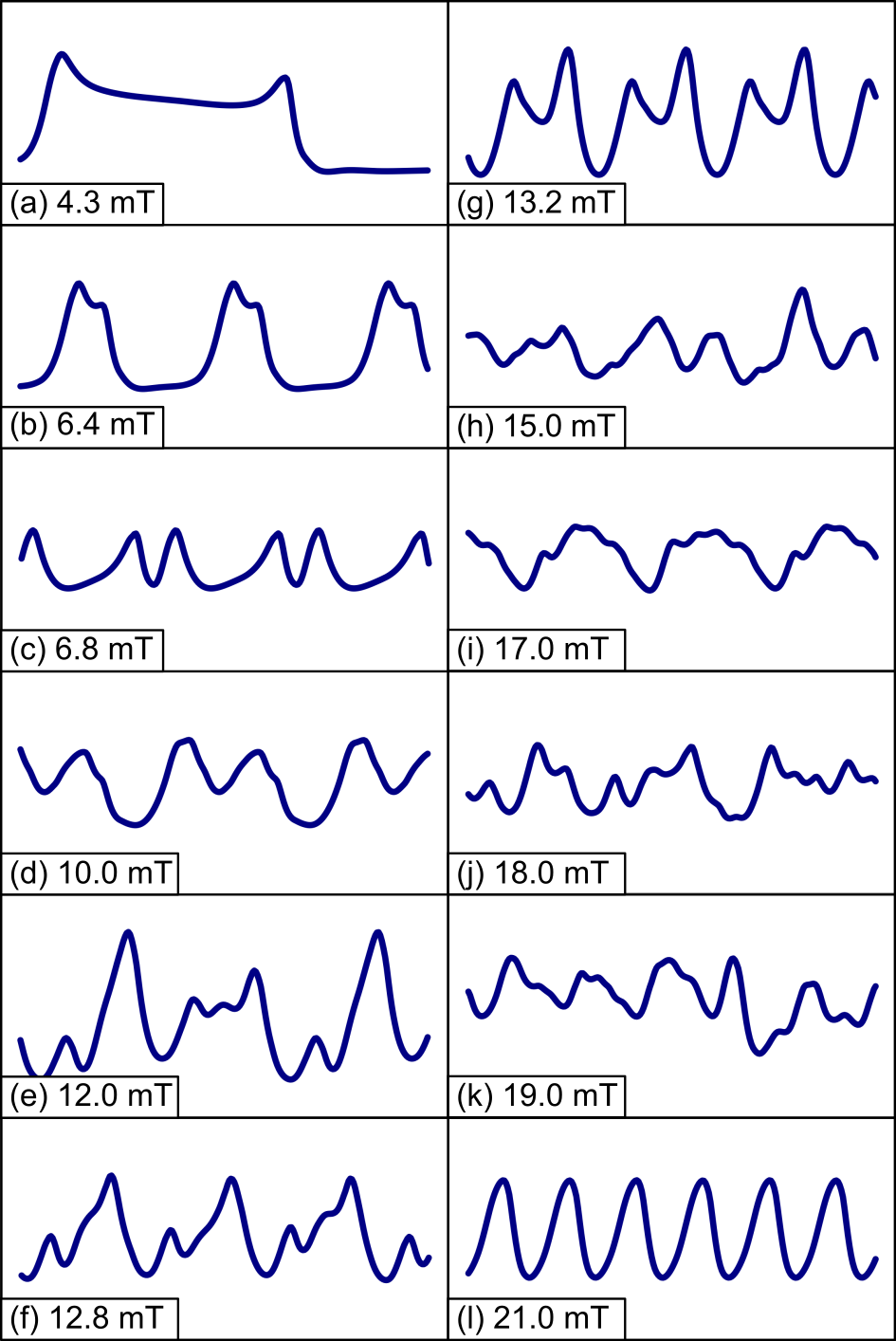} 
    \caption{Waveforms of the voltage $U(t)$ for the nanotube with $R=390$\,nm at $j_\mathrm{tr} = 16$\,GA/m$^2$ for a series of magnetic fields, as indicated in the figure. The scale of the $x$-axis is $125$\,ps while the scale of the $y$-axis is $30\,\mu$V. The corresponding vortex arrangements are shown in the respective panels in Fig.\,\ref{fig:trajectories}.}
    \label{fig:waveforms}
\end{figure}

Figure\,\ref{fig:waveforms} presents the waveforms $U(t)$ for the nanotube with $R=390$\,nm for a series of $B$ values, augmenting the essential features of the spectra in Fig.\,\ref{fig:spectra}(c). For instance, at $6.4$\,mT the peaks exhibit the periodicity with the main frequency. At $10$\,mT, one peak appears full-fledged, while the subsequent peak has a beveled shape. This behavior explains the presence of the subharmonics $f_{1/2}$. When the spectrum is blurry, for example at $18$\,mT, $U(t)$ manifests chaotic oscillations.

\begin{figure}
    \centering
    \includegraphics[width=1\linewidth]{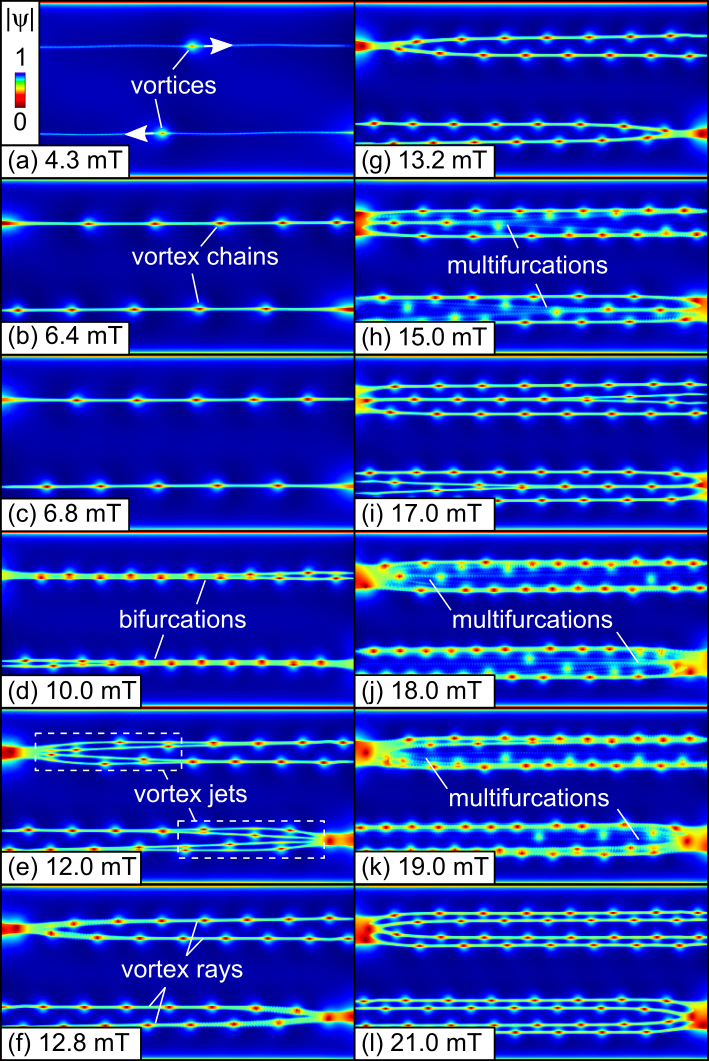} 
    \caption{Snapshots of the absolute value of the superconducting order parameter $|\psi|$ overlaid with the accumulated vortex paths for $j_\mathrm{tr} = 16$\,GA/m$^2$ at different values of the magnetic field, as indicated in the figure. At $B\lesssim8$\,mT vortices follow the same path within the half-tubes, forming single \emph{vortex chains}. At $8$\,mT$\lesssim B \lesssim 21$\,mT the vortex trajectories undergo multifurcations, giving rise to more complex tweezers-like patterns composed of \emph{vortex jets} consisting of two or more \emph{vortex chains}.}
    \label{fig:trajectories}
\end{figure}

To explore the factors contributing to the emergence of subharmonics, in Fig.\,\ref{fig:trajectories} we draw the accumulated vortex paths overlaid with snapshots of the absolute value of the superconducting order parameter $|\psi|$. As the field increases, starting from one and the same point of nucleation, vortices follow different paths\,\cite{Rez15rpj}. If there is only one path for the vortices in each half-tube (e.g., at $4.3$, $6.4$ and $6.8$\,mT), then there are no subharmonics in the spectrum. As a rule, when a subharmonic $f_{1/m}$ appears, it corresponds to the emergence of $m$ distinct paths. For example, there are two paths at $10$\, mT and four paths at $12$\,mT. Vortices follow these paths in an alternating order. At $21$\,mT, a four-path regime re-occurs. However, these four paths do not merge back into two paths, which differs from the regime observed at $12$\,mT. In some ranges of $B$ (e.g., at $15$, $18$, and $19$ mT), the paths do not exhibit well-defined patterns, displaying a blurred distribution. This behavior is reflected in the blurry spectra in Fig.\,\ref{fig:spectra}(c) and in the aperiodicity of the voltage.

In summary, the presence of subharmonics $f_q$ in the non-blurry spectrum can be attributed to the alternation of vortices between a number of paths in a periodic manner. In regimes with a blurry spectrum, vortices are not restricted to specific paths, though some paths may be more probable than others. The geometrical properties of the nanotube play a crucial role in determining a specific regime characterized by certain paths. While the number of vortex paths remains stable even with changes in the nanotube length, the radius of the nanotube has a significant influence on it. The dependence of the voltage spectrum and the vortex trajectories on the nanotube radius will be discussed in Sec.\,\ref{sec:confinement}.

\subsection{Voltage amplitude oscillations}
\label{sec:amp_osc}

Moving vortices induce a non-zero voltage between the slit banks. As vortices typically nucleate and denucleate at different instants of time, the number of vortices in the open tube is alternating. This may create appreciable oscillations of the induced voltage as a function of time $U(t)$. However, when the ratio $\tau = \Delta t_\mathrm{fl}/\Delta t_\mathrm{n}$ is an integer $n\in\mathbb{N}$, nucleation of vortices at one edge of the half-tube and their denucleation at the other edge of the half-tube occur nearly simultaneously. This leads to an almost constant number of vortices in the open tube, resulting in the suppression of the voltage oscillations. Such a discretization of the ratio $\tau$ can be achieved via magnetic induction and transport current variation.

\clearpage
To describe the amplitude of the alternating voltage phenomenologically within a simple model, we make use of the RMSD $\sigma_\mathrm{U}$ from the average voltage $\left<U\right>$
\begin{equation}
    \sigma_U =
    \sqrt{\left<
    \left(U - \left<U\right>\right)^2
    \right>} =
    \sqrt{\left<U^2\right> - \left<U\right>^2},
\end{equation}
where $\left<\ldots\right>$ means averaging over time. In our model, we assume that the motion of vortices is uniform and each moving vortex induces the same voltage $U_0$, i.e., the total voltage $U(t)=2U_0 N(t)$ is proportional to the total number of vortices $N(t)$ in each half-tube at the time instant $t$. The model does not account for the constant voltage component arising due to the normally conducting leads since this component does not contribute to the RMSD. Furthermore, we assume that there are $2N_0$ vortices in the open tube during the fraction of time $1-x$ and $2(N_0+1)$ vortices during the fraction of time $x$, where $x\in[0,1]$. Then, the average voltage and its RMSD read
\begin{subequations}
\begin{equation} \label{eq:avg.U}
    \left<U\right>
    =
    2 U_0 \left<N\right>
    =
    2U_0 (N_0 + x),
\end{equation}
\begin{equation}\label{eq:avg.deviations}
   \sigma_\mathrm{U} =
   2U_0\sqrt{\left<N^2\right>-\left<N\right>^2}=
   2U_0 \sqrt{x(1-x)}.
\end{equation}
\end{subequations}
The average number of vortices is determined by the ratio $\tau$, ${\left<N\right> = N_0 + x = \tau}$, leading to
\begin{equation}\label{eq:n_and_x}
    \left\lfloor \tau \right\rfloor = N_0,
    \qquad
    \left\{ \tau \right\} = x,
\end{equation}
and
\begin{equation}\label{eq:avg.deviations_2}
   \sigma_\mathrm{U} = 2U_0 \sqrt{\left\{ \tau \right\}\left(1-\left\{ \tau \right\}\right)},
\end{equation}
where $\lfloor\cdot\rfloor$ and $\{\cdot\}$ denote the integer and fraction parts of a number. A higher magnetic field favors a larger $\tau$, resulting in the oscillations of $\sigma_\mathrm{U}$ as a function of the magnetic field due to the periodicity of the function $\left\{ \tau \right\}$, namely, $\left\{ \tau + 1 \right\} = \left\{ \tau \right\}$.

The validity of this reasoning is confirmed by the RMSD $\sigma_\mathrm{U}$ shown in Fig.\,\ref{fig:spectra}(b) for $j_\mathrm{tr} = 16$\,GA/m$^2$. The oscillations of $\sigma_\mathrm{U}$ appear at $4.4$\,mT when vortex nucleation commences. As predicted above, these oscillations exhibit a clear periodicity. The oscillations period found from the simulations is $\Delta B \approx 0.64$\,mT. This implies that the ratio $\tau = \Delta t_\mathrm{fl} / \Delta t_\mathrm{n}$ is indeed a linear function of $B$ in the range of fields where the periodicity is accurate. For instance, the maximal and minimal values of $\sigma_\mathrm{U}$ are at $6.4$ and $6.8$\,mT, respectively. In the case of maximal $\sigma_\mathrm{U}$, the voltage waveform features a single pronounced peak per period [see, e.g., Fig.\,\ref{fig:waveforms}(b)], while in the case of minimal $\sigma_U$ [see, e.g., Fig.\,\ref{fig:waveforms}(c)] there are two peaks per period with twice smaller oscillations of the voltage. Nevertheless, the paths corresponding to the maximal and minimal $\sigma_\mathrm{U}$ are identical, forming straight paraxial lines in Fig.\,\ref{fig:trajectories}(b,c).

\begin{figure}
    \centering
    \includegraphics[width=\linewidth]{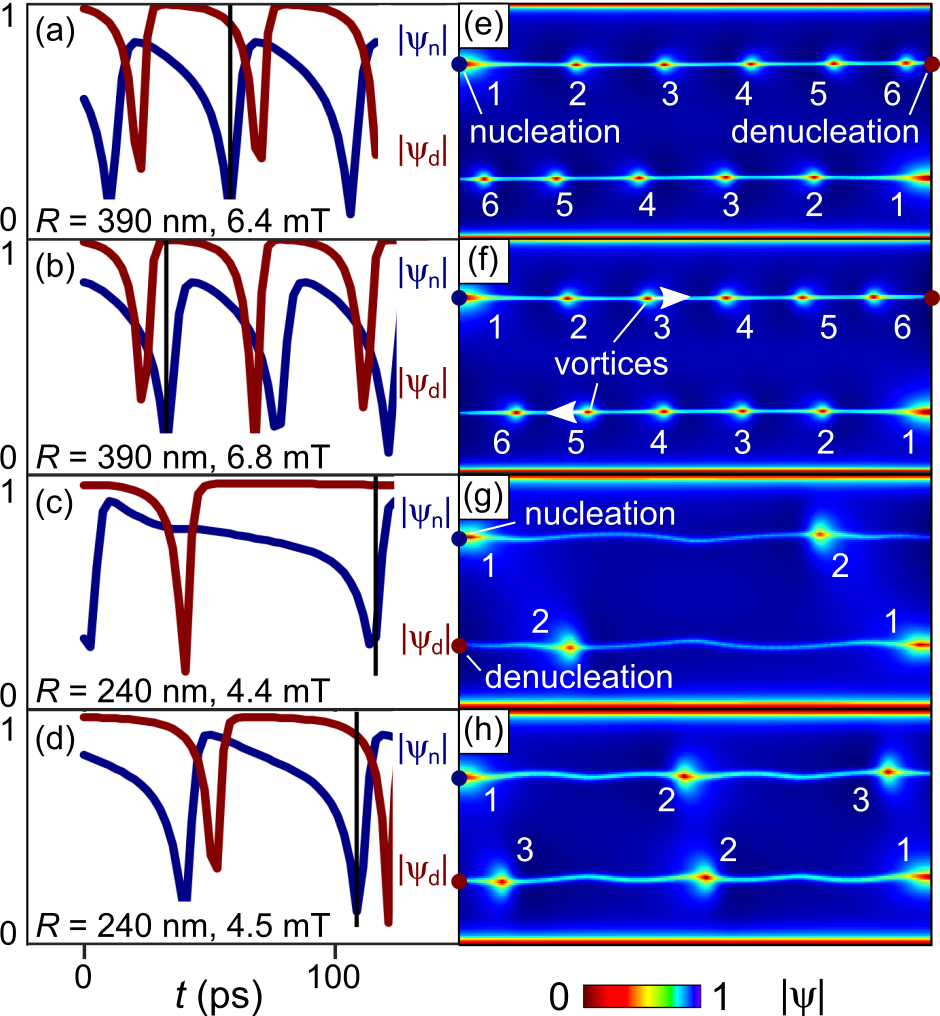}
    \caption{Evolution of $|\psi|$ at the points of the open nanotubes with radii $R = 390$\,nm (a,b,e,f) and $240$\,nm (c,d,g,h), where vortices nucleate, $\psi_\mathrm{n}$, and denucleate, $\psi_\mathrm{d}$. (e-h) Snapshots of the spatial dependence of $|\psi|$ in the steady regime. Panels (a,e) and (b,f) correspond to the maximum and minimum values of $\sigma_\mathrm{U}$, respectively. Panels (c,g) and (d,h) correspond to the regime before and after the jump, respectively. The light blue curves connecting the vortex cores in panels (e-h) indicate the accumulated vortex paths.}
    \label{fig:psi}
\end{figure}

The underlying mechanism of this effect, which involves the shift between the nucleation and denucleation events, is elucidated by Fig.\,\ref{fig:psi}. Figure\,\ref{fig:psi}(a-d) illustrates the evolution of the magnitude of the superconducting order parameter $|\psi|$ at the points, where vortex nucleation and denucleation occur. Sharp drops in the respective functions point to the nucleation or denucleation events. At $6.4$\,mT, when $\sigma_\mathrm{U}$ is close to its maximal value, the denucleation at the other edge of the same half-tube occurs when $|\psi_{n}|$ reaches its maximal value after the preceding nucleation. At $6.8$\,mT, when $\sigma_\mathrm{U}$ is close to its minimal value, nucleation occurs just after the denucleation at the other edge of the same half-tube. Minor variations between the peak shapes stem from the rather coarse time steps. Visualization of the order parameter distribution in Fig.\,\ref{fig:psi}(e,f) further corroborates the impact of the time shift between nucleation at one edge and denucleation at the other edge of the half-tube on the extrema of $\sigma_\mathrm{U}$.

\section{Discussion}
We first discuss the evolution of the voltage frequency spectrum $U_\mathrm{f}$ as a function of the magnetic field, with the aim to link the changes in $U_\mathrm{f}$ to transitions between various vortex array arrangements. We next proceed to confinement effects in nanotubes of smaller radii and compare the vortex arrangements with those for planar thin films. The discussion is concluded with remarks on the applicability of the model and its potential experimental examination.

\subsection{Voltage spectra and non-diverging vortex jets}
\label{sec:jets}
The major findings of the present work are related to two groups of results. First, predictions are made for the voltage frequency spectrum which evolves between an $nf_1$-type and an $\frac{m}{n}f_1$-type [$f_1$: vortex nucleation frequency], and is blurred in certain ranges of magnetic fields and currents. The frequency voltage spectra can be measured experimentally by connecting a nanotube to a spectrum analyzer\,\cite{Dob20pra} or placing a microwave antenna in the vicinity of the sample\,\cite{Gol94prb,Dob18nac}.

Second, predictions are made for novel vortex array arrangements and transitions between them with increase of $B$ and $j_\mathrm{tr}$. In contradistinction to planar thin films, in which vortices nucleate at \emph{many} points along the sample edge\,\cite{Bud22pra}, the geometry-induced non-uniformity of the magnetic field in the nanotube makes the vortices to nucleate only at the \emph{two} edge points where $B_\mathrm{n}$ is close to maximum. Imaging of vortex arrays represents a challenging task requiring a local-probe technique, which should be fast and sensitive enough to resolve the fast motion of individual vortices. While a number of methods, including scanning tunneling microscopy\,\cite{Tro99nat}, magnetic force microscopy\,\cite{Aus09nph}, scanning superconducting quantum interference device (SQUID) microscopy\,\cite{Kre16nal}, and scanning Hall
probes\,\cite{Col06nam,Sil10prl} were employed to image slowly moving vortex arrays, only scanning SQUID-on-tip microscopy\,\cite{Emb17nac} could resolve the properties of high-speed vortices in planar thin films so far, and we are not aware of any similar studies for 3D manifolds until now. The features in the global observables, such as voltage and its frequency spectrum, hence contain essential information on the arrangements of vortices, which is hard to acquire directly and which can be analyzed with the aid of modeling.

At low magnetic fields ($4$-$8$\,mT in Fig.\,\ref{fig:trajectories}), vortices form vortex chains and, distinct from planar films where vortex chains evolve into diverging jets because of repulsive vortex-vortex interaction, vortex chains in open nanotubes of the studied length remain stable for up to about ten vortices [see Figs.\,\ref{fig:trajectories}(a-c) and\,\ref{fig:psi}(e-h)], due to the \emph{constraint effect} of the non-uniform magnetic field. This differs dramatically from both, the theoretical predictions for planar thin films\,\cite{Bez22prb,Ust23etp,Bev23pra} and the experimental observations for planar constrictions\,\cite{Emb17nac}.

At moderately strong fields ($8$-$13$\,mT in Fig.\,\ref{fig:trajectories}), a transition to a vortex-jet regime occurs. However, in contradistinction to diverging vortex jets in planar structures\,\cite{Emb17nac,Bez22prb,Ust23etp,Bev23pra}, after the initial bi- or multifurcation of the vortex paths, the vortex jets in the half-tubes remain \emph{non-diverging} as the vortex trajectories tend to further run parallel to the nanotube axis. Again, the vortex jet non-divergence stems from the constraint of vortex arrays to the tube areas close to maximum $B_\mathrm{n}$.

At higher magnetic fields ($13$-$21$\,mT in Fig.\,\ref{fig:trajectories}), further transitions from a two-chain vortex jet to three- and four-chain vortex jets occur. These transitions are mediated by multifurcations of the vortex trajectories (such as at $15$ and $18$\,mT in Fig.\,\ref{fig:trajectories}), decisively affecting the voltage frequency spectrum.

As a major result of comparison of the voltage spectra in Fig.\,\ref{fig:spectra} with the vortex arrangements in Fig.\,\ref{fig:trajectories}, we conclude that $nf_1$-type voltage spectrum corresponds to a vortex chain, $\frac{n}{m}f_1$-type voltage spectrum points to a vortex jet consisting of $m$ vortex chains, and a blurry voltage spectrum emerges in consequence of chaotic multifurcations of the vortex trajectories.

To summarize, the major features of vortex jets in open tubes are the following: (i) a vortex chain remains stable for up to about ten vortices, (ii) vortex jets may consist of more than two vortex chains and (iii) these jets are not diverging, as distinct from the previously studied planar geometries\,\cite{Emb17nac,Bez22prb}. The constraint of the vortex motion within the half-tubes, which becomes apparent as a suppressed vortex chain--vortex jet transition for three vortices and the non-divergence of vortex jets, are attributed to the pronounced non-uniformity of the magnetic field. Since the field non-unifromity is enhanced when reducing the nanotube radius, the confinement effects should be stronger for tubes of smaller radii, as will be corroborated in Sec.\,\ref{sec:confinement}.

\subsection{Microwave generation due to moving fluxons}
\label{sec:generation}

An array of moving vortices is known to generate alternating electromagnetic field\,\cite{Kul66spj,Bul06prl,Dob18apl,Dob18nac}. Since the instrumentation required for the detection of generation depends on the frequency range, the quantitative predictions of frequencies in our simulations should be treated with particular attention.

On the one hand, major features in the voltage spectra are predicted for the microwave range. The frequencies $10$-$100$\,GHz correspond to the voltage oscillation periods between $100$\,ps to $10$\,ps, respectively, see Fig.\,\ref{fig:waveforms}. On the other hand, these frequencies are by two orders of magnitude higher than one could expect for Nb films, where the vortex dynamics is limited by $\sim1$\,km/s vortex velocities\,\cite{Bez19prb}. At higher velocities, the vortex cores collapse because of the escape of unpaired electrons from the vortex cores, resulting in an abrupt onset of a highly resistive state. This phenomenon is termed flux-flow instability\,\cite{Lar86inb,Dob23inb}. For an array of ten vortices moving at $v\simeq1$\,km/s under a rather large transport current, the frequency associated with the vortex (de)nucleation can be estimated as $f = 10 v/L$, where $L=5\,\mu$m is the nanotube length, yielding $f= 2$\,GHz.

The fast motion of vortices in a chain also implies that the unpaired electrons (quasiparticles), which escape from a vortex core, have enough time to relax before being trapped by the subsequent vortex, requiring characteristic times $\tau$ larger than the time of quasiparticle energy relaxation, $\tau > \tau_\varepsilon = L/(10v) = 500$\,ps. This difference in the time scales stems from the TDGL equation involving a much faster dynamics of the superconducting order parameter rather than capturing a particular quasiparticle relaxation mechanism for a given superconductor. In the numerical modeling, the adequate frequency and time scales for the vortex dynamics can be obtained by adding a numerical factor in front of the order parameter derivative with respect to time. After this adaptation of the diffusion coefficient, the features in $U_\mathrm{f}$ fall into the radiofrequency and low-frequency microwave range $0.1$-$10$\,GHz, which are relevant for communication and information processing.

\subsection{Confinement of vortex motion}
\label{sec:confinement}
As demonstrated above, the vortex dynamics in nanotubes of $390$\,nm radius exhibit a diverse range of behaviors. The system undergoes distinct transitions in its operating mode, which are reflected in the voltage spectra and vortex paths. In order to extend operations within a single regime over a wide range of magnetic fields, one can force vortices to move one behind another following a single path. This can be accomplished by reducing the nanotube radius, as it is illustrated for a nanotube with $R=240$\,nm. In the case of this smaller nanotube, we subject it to a larger $j_\mathrm{tr} = 20$\,GA/m$^2$, since the commencement of vortex nucleation necessitates a larger current or higher magnetic field compared to the bigger nanotube.

\begin{figure}
    \centering
       \includegraphics[width=0.98\linewidth]{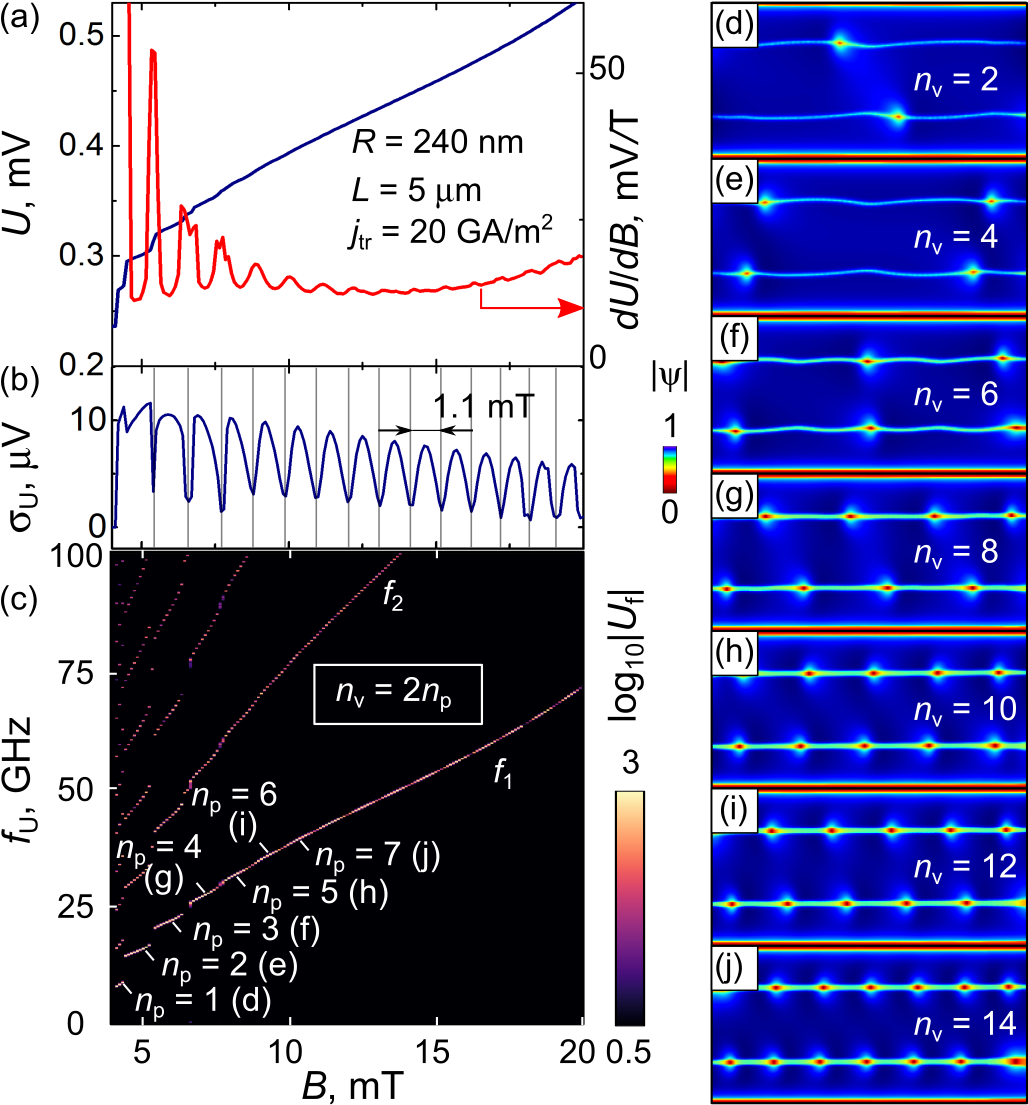}
    \caption{(a-c) Voltage as a function of the magnetic field $B$ for a nanotube with $R=240$\,nm at $j_\mathrm{tr} = 20$\,GA/m$^2$. From top to bottom: average voltage $U$, RMSD $\sigma_\mathrm{U}$ of the voltage, and the voltage spectrum $U_\mathrm{f}$. (d-f) Snapshots of the absolute value of the superconducting order parameter $|\psi|$ for a series of magnetic field values, as indicated in the figure. The number of vortices in the tube is equal to the doubled consecutive number of the frequency plateau, $n_v = 2 n_\mathrm{p}$.}
    \label{fig:240nm}
\end{figure}

The voltage spectrum for the nanotube with $R=240$\,nm is shown in Fig.\,\ref{fig:240nm}(a-c). The spectrum exhibits a $nf_1$-behavior arising from a single vortex path in each half-tube throughout the entire range of magnetic fields, free of blurriness and subharmonics. The oscillations of $\sigma_\mathrm{U}$ are more pronounced and extend over a greater number of periods. Furthermore, a new effect is observed, characterized by jumps in the main frequency (i.e., the nucleation frequency) at each minimum of $\sigma_\mathrm{U}$, see Fig.\,\ref{fig:240nm}(a-c). As the nucleation frequency affects the number of vortices at a given time instant in the tube, this leads to a small jump in the average voltage $U$ as well. Notably, frequency jumps occur over a small interval of $B$ rather than at a single $B$ value. In such intervals, all odd harmonics, including $f_1$, vanish. Typically, the nucleation processes are phase-synchronized in both half-tubes. However, when odd harmonics are absent, alternate vortex nucleations in the half-tubes occur, resulting in an anti-phase synchronization.

It is worth noting that, in the general case, the actual number of vortices $n_\mathrm{v}$ in the superconductor can be larger (if there are edge defects\,\cite{Bev23pra}) or smaller (for perfect edge barriers\,\cite{Bud22pra}) than follows from the estimate $n_\mathrm{v}\Phi_0 = BS$\,\cite{Bev23pra} [$n_\mathrm{v} = 1, 2, \dots$, $\Phi_0$: magnetic flux quantum; $S$: area of the relevant sample surface] and this complicates the deduction of the vortex velocity from the measured voltage. The reason for this is that a larger number of slow-moving fluxons can induce the same voltage as a small number of fast-moving ones\,\cite{Bev23pra}. The plateau number $n_\mathrm{p}$ between two consecutive jumps in the frequency of the microwave radiation can therefore be suggested for the deduction of the number of fluxons $n_\mathrm{v}$ moving in tubes at low magnetic fields. However, distinct from planar thin films, where the serial number, $n_\mathrm{k}$, of a kink in the differential resistance corresponds to the number of vortices crossing the constriction, $n_\mathrm{v}=n_\mathrm{k}$, the number of vortices in 3D open tubes is given by $n_v = 2 n_\mathrm{p}$, because of the presence of two half-tubes.

The hypothesis of a stronger intervortex interaction in nanotubes of smaller radii is supported by the vortex paths in Fig.\,\ref{fig:psi}(c,d,g,h). Namely, at $4.4$\,mT, before the first jump of $f_\mathrm{U}$, nucleation at an edge of one half-tube takes place when a vortex in the other half-tube is positioned \emph{far} from the same edge. A slight increase in the magnetic field to $4.5$\,mT leads to a transition to a different regime, where nucleation at an edge of one half-tube occurs when a vortex in the other half-tube is \emph{close} to the edge where its denucleation takes place. It is assumed that a vortex which is close to its denucleation promotes a vortex nucleation at the same edge of the other half-tube due to their attraction. The accumulated paths of vortices are no longer straight lines. Instead, they exhibit slight deviations from straight lines, implying an appreciable intervortex interaction between the vortices moving in the opposite half-tubes. A reference simulation for a single half-tube of the same radius is shown in Fig.\,\ref{fig:planar}(a-c). It does not reveal any jumps in $f_\mathrm{U}$, thus supporting the role of the interaction between vortices moving in the opposite half-tubes.

\subsection{Comparison with planar structures}
\label{sec:planar}

The considered vortex arrangements in open nanotubes can be compared with those in planar systems studied in previous works\,\cite{Emb17nac,Bez22prb}.
For instance, for planar thin film Pb constrictions with a strongly non-uniform current density distribution, the vortices nucleated at one edge of its narrowest part and were driven by noncollinear local Lorentz-type forces because of the strong bending of the current streamlines in the constriction. With increase of the flux density and transport current, this led to multifurcations of the vortex trajectories, while the intervortex repulsion was not too strong because of the small penetration depth $\lambda\sim50$\,nm for Pb thin films.

In a different study\,\cite{Bez22prb} for planar thin MoSi microstrips at zero magnetic field, a small narrow defect at the microstrip edge was used as a point for vortex nucleation because of the local suppression of the edge barrier by the transport current. While the distribution of the current density far away from the defect was uniform, the long-ranged intervortex repulsion ($\lambda\sim500$\,nm for MoSi) led to the fact that in the presence of fluctuations or material inhomogeneity the vortex chain regime manifested for only up to two vortices, while for three and more vortices vortex jets emerged. With a further increase of the transport current, the vortex velocity component across the microstrip commensed to significantly exceed the velocity component along the microstrip, leading to a \emph{descrease} of the jet opening angle. At yet stronger currents, the unpaired electrons left behind the vortex cores commenced to attract strongly the other vortices, resulting in the formation of a ``vortex river'' (chain of vortices with depleted cores and equivalent to a dynamically formed Josephson junction or a phase-slip line). Since the TDGL equation in our work was solved without the thermal balance equation, we restricted our consideration by not-too-strong currents (depairing current density can be estimated as $23.1$\,GA/m$^2$). Qualitatively, however, one can expect the formation of vortex rivers (if heat removal is fast enough) or normally conducting domains (if heat removal is insufficient) at yet stronger currents.

\begin{figure}
    \centering
        \includegraphics[width=0.98\linewidth]{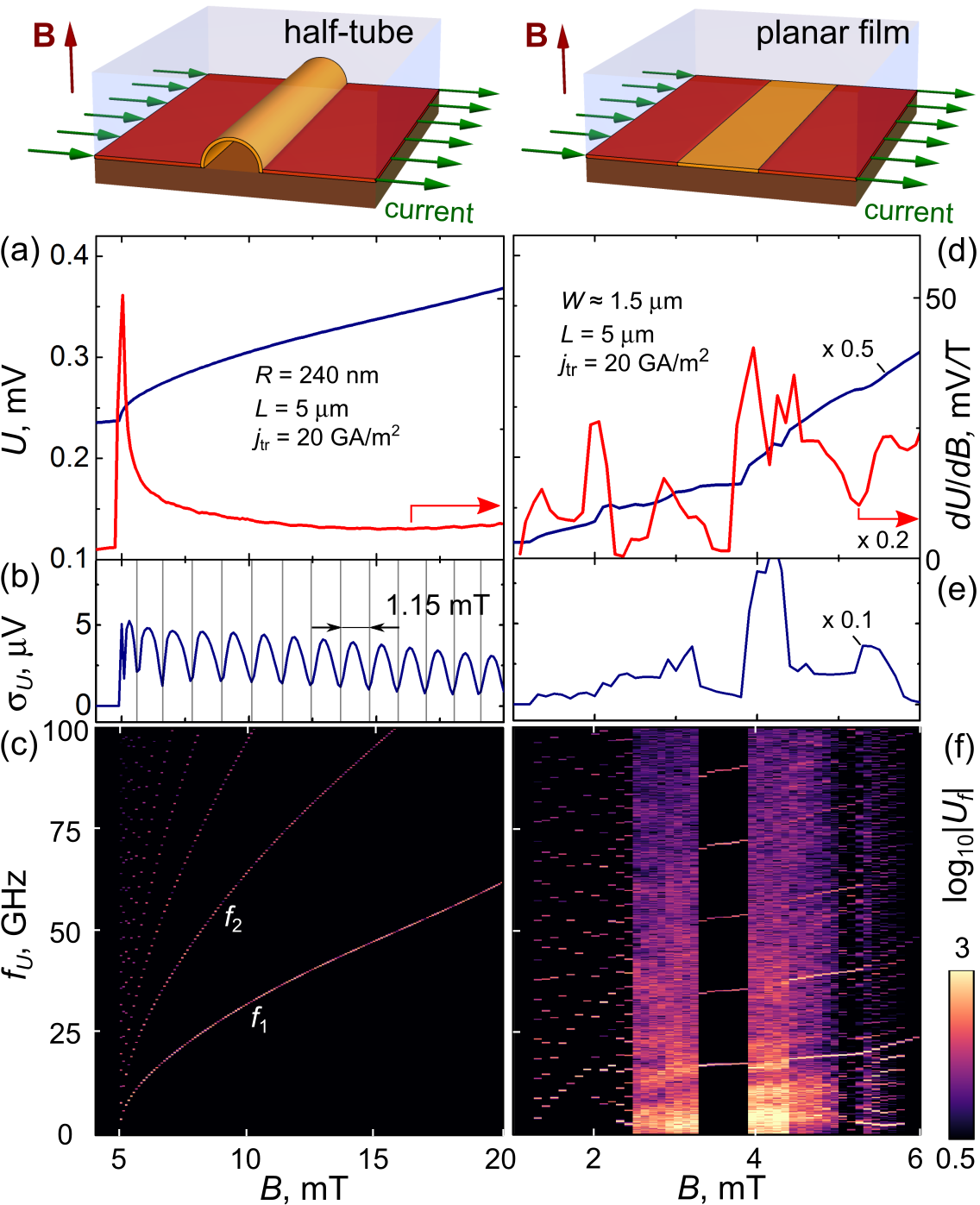}
    \caption{Voltage as a function of the magnetic field $B$ for half-tube with $R = 240$\,nm (a-c) and a planar structure (d-f) with width $W=2\pi\times240\,$nm at $j_\mathrm{tr} = 20$\,GA/m$^2$. From top to bottom: average voltage $U$, RMSD $\sigma_\mathrm{U}$ of the voltage and the voltage spectrum $f_\mathrm{U}$.}
    \label{fig:planar}
\end{figure}

For comparison, Fig.\,\ref{fig:planar}(d-f) presents the average voltage, its RMSD and frequency spectrum as a function of the magnetic field for a planar membrane with the width $W = 1.5\,\mu$m $\approx 2\pi \times 240\,$nm and the length $L = 5\,\mu$m. In contrast to the $nf_1$ voltage spectra for the nanotube with $R =240$\,nm in Fig.\,\ref{fig:240nm}(c) and the half-tube in Fig.\,\ref{fig:planar}(c), $U_\mathrm{f}(B)$ for the planar structure contains subharmonics and is blurried in certain ranges of currents and fields. These features are attributed to the fact that in the absence of edge defects, the nucleation of vortices occurs at different points along the entire edge of the structure and various vortex arrangements as, e.g., in Fig.\,5 of Ref.\,\cite{Dob20nac}. Obviously, these arrangements are very distinct from the vortex patterns revealed for open nanotubes in the present work and they are qualitatively similar to those for the previously studied planar wide MoSi strips\,\cite{Bud22pra}. As additional point, superconductivity in the planar structure is destroyed at a smaller field ($6$\,mT) in comparison with the open tube and the half-tube. This finding is attributed to the presence of areas with smaller $B_\mathrm{n}$, which can be viewed as an effective reduction of the external magnetic field in 3D structures.

We note that for planar superconductor strips, a higher-correlated regime in the vortex dynamics can be realized via creation of a slit for the vortex nucleation at a given point\,\cite{Bez22prb}. The slit locally reduces the strip cross-section\,\cite{Bud22pra} and leads to an increase of the current density (current-crowding effect)\,\cite{Cle11prb}. By contrast, the correlated dynamics of vortices in open nanotubes are realized via the non-uniformity of $B_\mathrm{n}$ rather than current-crowding effects. At the same time, while the extension of a planar structure into the third dimension makes the vortices to nucleate at a given edge point close to maximum $B_\mathrm{n}$, our simulations for a half-tube have not revealed peaks in the average voltage as a function of current and field [Fig.\,\ref{fig:planar}(a)]. This is distinct from differential resistance oscillations in slitted planar constrictions upon increase of the fluxon number by one  [see Figs.\,2 and 3 in Ref.\,\cite{Ust23etp} and Figs.\,2 and 3 in Ref.\,\cite{Bev23pra}]. However, in our model, peaks in $U(B)$ and jumps in the microwave generation frequency are predicted for nanotubes of rather small radii [Fig.\,\ref{fig:240nm}(a)], pointing to the decisive role of the interaction of vortices in the both half-tubes for the correlated vortex dynamics.

\subsection{Applicability of the model}
\label{sec:applic}
An experimental examination of our predictions requires an outline of the effects which are captured and which are not captured by the considered model.

First, as detailed in the Appendix, we consider a steady conversion of the transport current entering the normal leads into the superconducting current inside the tube over a certain transition area embedding the scalar potential. This is achieved by a self-consistent solution of the TDGL equation and the Poisson equation for the scalar potential. However, the model does not account for further areas with suppressed superconductivity which could appear because of sharp bends of the membrane in the regions where it touches the substrate. Like for the aforementioned transition area from normal metal to superconductor, an account for the suppression of superconductivity at the slit banks is expected to lead to a constant voltage contribution in addition to the voltage discussed in the present work.

Second, a uniform temperature distribution over the tube is assumed, what requires an efficient Joule heat removal from the nanotube to the cryogenic environment\,\cite{Bez19prb}. Experimentally, this can be achieved, e.g., via embedding the nanotube into some insulating material, as indicated by the semitransparent box in Fig.\,\ref{fig:scheme}(a), which is a good heat conductor at low temperatures. For an account for a non-uniform temperature distribution, the TDGL equation should be complemented with the thermal balance equations for the electron and phonon temperatures\,\cite{Bud22pra}, what makes the numerical solution much more time-consuming even for planar thin films.

Third, the interaction between vortices in the upper and lower half-tubes via the stray fields is neglected in the model, since this interaction has been proven to be small for tubes of the considered radii and thicknesses\,\cite{Smirnova20,Bog22prb}. Thus, vortices interact with each other via screening currents only. This approximation is justified for films with thicknesses $d\leq50$\,nm and radii $R\geq 200$\,nm. Qualitatively, an account for the interaction of vortices via stray fields is expected to lead to a mutual breaking of vortices moving in the both half-tubes\,\cite{Bog22prb}.

Fourth, the geometry-induced non-uniformity of $B_\mathrm{n}$ weakens requirements to the edge quality of the nanotube\,\cite{Bud22pra}. This is because of vortex (de)nucleation occurring not along the entire free edges, but in the regions close to $B_\mathrm{n}$ maximum. Thus, even though the vortex jets in the different half-tubes interact with one another via the screening currents, they remain constrained to the regions where $B_\mathrm{n}$ is close to maximum.

Finally, vortex pinning and edge defects are not included in the model. This approximation is hence justified for materials characterized by weak volume pinning, such as amorphous superconducting films. At the same time, the roll-up technology\,\cite{Thurmer08,Thurmer10,Loe19acs}, which is based on the release of mechanical strain in the substrate materials, implies the occurrence of defects in superconductor membranes. Within the framework of the considered model, vortex pinning effects on the induced voltages and their spectra are discussed elsewhere\,\cite{Bra24arx}.

\section{Conclusion}
In summary, we have considered the dynamics of vortices in open superconductor nanotubes exposed to an azimuthal transport current and a perpendicular-to-tube-axis magnetic field. On the basis of a numerical solution of the TDGL equation, the induced voltage $U$ oscillations at GHz-frequency have been predicted with spectra evolving between $nf_1$-type and $\frac{n}{m}f_1$-type and blurred in certain ranges of currents and fields. Snapshots of the spatiotemporal distributions of the superconducting order parameter have revealed that an $nf_1$-spectrum corresponds to a single vortex-chain regime typical for low magnetic fields and for tubes of small radii. At higher fields, an $\frac{n}{m}f_1$-spectrum implies the presence of $m$ vortex chains in the vortex jets which, in contrast to planar thin films, are not diverging because of constraint to the tube areas where $\mathbf{B}_\mathrm{n}$ is close to maximum. A blurry spectrum points to complex arrangements of vortices emerging because of multifurcations of their trajectories. A reduction of the nanotube radius results in promotion of single-vortex-chain regimes and noticeable deformations of vortex paths due to the enhanced interaction between vortices in the both half-tubes. In addition, due to a stronger confinement of single vortex chains in tubes of small radii, peaks in $dU/dB$ and jumps in the frequency of microwave generation occur when the number of fluxons moving in the half-tubes increases by one.

Our findings suggest that vortex jets can be constrained and possibly steered using the curvature of 3D superconductor membranes. Namely, the presence of a convexity with a maximum of $\mathbf{B}_\mathrm{n}$ may act as a conveyor for vortices, offering a new paradigm for the vortex guiding\,\cite{Dob20pra}. A smaller curvature radius may be used then for focusing of a vortex jet and extending the vortex chain regime to stronger currents and higher magnetic fields. An interesting regime could be expected for rather small curvatures of the superconducting membranes, where the divergence of vortex jets peculiar to 2D planar thin films competes with the trend towards the formation of paraxial vortex chains (jet focusing) in a 3D membrane.

Overall, while time- and space-resolved experimental studies of vortex dynamics in supeconductor 3D nanoarchitectures remain to be performed, deduction of vortex configurations from an analysis of global observables -- as discussed in the present work -- represents one of the viable approaches to that end. Our predictions regarding the average voltage, voltage oscillations and voltage frequency spectra can be examined experimentally for, e.g. open Nb nanotubes fabricated by the self-rolling technology. In all, our findings are essential for novel 3D superconductor devices which can operate in few- and multi-fluxon regimes.

\begin{acknowledgments}
The authors are grateful to Edward Goldobin for fruitful discussions. IB is grateful to Victor Ciobu for technical support, including the generous provision of several servers that were instrumental for the calculations. OVD acknowledges the Austrian Science Fund (FWF) for support through Grant No. I 6079 (FluMag). VMF expresses his thanks to the European Cooperation in Science and Technology for support via Grant E-COST-GRANT-CA21144-d8436ac6-b039a83c-fa29-11ed-9946-0a58a9feac02 and to the ZIH TU Dresden for providing its facilities for high throughput calculations. This article is based upon work supported by the E-COST via Action CA21144 (SuperQuMap).
\end{acknowledgments}

\appendix
\section{Numerical modeling}
The open nanotube in a homogeneous perpendicular-to-tube-axis magnetic field was mapped to a planar membrane in a modulated out-of-plane field, see Fig.\,\ref{fig:scheme}. The mathematical model of the nanotube is represented by a 2D surface denoted as $D$, parameterized by orthonormal coordinates $x$ and $y$, along with a normal unit vector $\mathbf{n}$. Surface $D$ is embedded in 3D space with Cartesian coordinates $X$, $Y$, and $Z$. The modeling was based on a numerical solution of the 2D TDGL equation, which, in its dimensionless form, reads
\begin{equation}
\label{eq:gl}
    (\partial_t + i \varphi) \psi = \left(\boldsymbol{\nabla} - i \mathbf{A}\right)^2 \psi + (1 - |\psi|^2)\psi.
\end{equation}
Here, $\partial_t$ denotes the derivative with respect to time $t$, $\boldsymbol{\nabla}$ is a 2D nabla operator on the surface, $\mathbf{A}$ is the vector potential, the scalar potential $\varphi$ determines the electric field $\mathbf{E}=-\boldsymbol{\nabla}\varphi$, and $\psi\equiv \psi(x,y,t)$ is the complex superconducting order parameter, which depends on the coordinates $x$ and $y$ and evolves with time $t$. The gauge of the vector potential $\mathbf{A}$ is chosen such that the following expressions are zero at the surface of the nanotube $D$: $\nabla_\mathbf{n} A_\mathbf{n} = A_\mathbf{n} = 0$, where subscript $\mathbf{n}$ denotes the projection of the corresponding vector onto the normal vector $\mathbf{n}$. Being defined in the entire 3D space, the vector potential determines the magnetic induction $\mathbf{B}=[\boldsymbol{\nabla}_{3D} \times \mathbf{A}]$, where $\boldsymbol{\nabla}_{3D}$ is the nabla operator in the 3D space. The units for the dimensionless quantities in Eqs.\,(\ref{eq:gl})-(\ref{eq:phi_bc}) are provided in Table\,\ref{tab:scale_1}.

The superconducting current density is determined as ${\mathbf{j}_\mathrm{sc} = \mathfrak{Im}(\psi^* (\boldsymbol{\nabla} - i \mathbf{A}) \psi)}$.
The effects of the magnetic field induced by the superconducting currents are neglected. The applicability of this approximation is discussed elsewhere\,\cite{Smirnova20,Bog22prb}.

The Poisson equation for the scalar potential $\varphi$ follows from the continuity of the total current density which is given by the sum of the superconducting $\mathbf{j}_\mathrm{sc}$ and normal $\mathbf{j}_\mathrm{n}$ components
\begin{equation} \label{eq:poisson}
    \Delta \varphi = \frac{1}{\sigma} \boldsymbol{\nabla} \cdot \mathbf{j}_\mathrm{sc},\qquad
    \mathbf{j}_\mathrm{n} = -\sigma\boldsymbol{\nabla}\varphi.
\end{equation}

The boundary conditions\,\cite{thinkham1996} for the TDGL read
\begin{equation}
    \left. \left( \partial_y - i A_y \right)\psi \right|_{\partial D_y}= 0,\qquad
    \left.\psi\right|_{\partial D_x} = 0,
\end{equation}
where $\partial D_x$ and $\partial D_y$ are the boundaries corresponding to the ends of the intervals for $x$ and $y$, respectively, see Fig.\,\ref{fig:scheme}(c). The transport current density $\mathbf{j}_\mathrm{tr}=j_\mathrm{tr} \mathbf{e}_x$ is introduced through the boundary conditions for the scalar potential
\begin{equation}\label{eq:phi_bc}
    \left. \partial_y \varphi \right|_{\partial D_y} = 0,\qquad
    \left. \partial_x \varphi \right|_{\partial D_x}  = - j_\mathrm{tr}/\sigma.
\end{equation}

The scalar potential is split into two terms, namely the non-divergent potential $\varphi_\mathrm{ndiv}$ and the induced potential $\varphi_\mathrm{ind}$. This separation allows for a faster convergence of the numerical algorithm\,\cite{Bog22prb,Sad15jcp} used for solving the Poisson equation.

As gauge-dependent numerical schemes may introduce enormous errors\,\cite{Kat93prb}, the link variables\,\cite{Rez20cph,Bog22prb} were used for both, the vector potential $\mathbf{A}$ (conjugated with coordinates) and the scalar potential $\varphi$ (conjugated with time). The voltage between the leads was calculated as a difference of the scalar potentials averaged over the lead length
\begin{equation}
    U(t) = \frac{1}{L} \int_0^L dy \left( \varphi(t, W, y) - \varphi(t, 0, y) \right).
\end{equation}

The modeling was done for parameters typical for Nb structures, see Table\,\ref{tab:params_1}. For the nanotube with $R=390$\,nm, a grid of $192\times384$ ($x\times y$) points and a time step $\Delta t = 0.025$\,ps were used. For other sizes and structures, the number of points was chosen to result in approximately the same density of the grid points per unit length. An iterative method of solving the Poisson equation was used until the absolute value of the difference between the left and right sides of Eq.\,(\ref{eq:poisson}) became smaller than $0.004$ for all grid points.

More details on the applicability of the mathematical model, link variables, splitting of the electric potential, and further numerical details are given elsewhere\,\cite{Rez20cph,Bog22prb}.

\begin{table}[b!]
\caption{Dimensional units for quantities in in Eqs.\,(\ref{eq:gl})--(\ref{eq:phi_bc}).
\label{tab:scale_1}}
\begin{center}\begin{tabular}{|l|c|c| }
 \hline
 Parameter & Unit & Value for Nb at $T/T_c = 0.952$ \\
 \hline
 Time & $\xi^2/D$ & 2.8 ps
 \\
 Length & $\xi$ & 60 nm
 \\
 Magnetic field & $\Phi_0 / 2\pi\xi^2$ & 92 mT
 \\
 Current density & $\hslash c^2 / 8\pi\lambda^2 \xi e$ & 60 GA $\text{m}^{-2}$
 \\
 Electric potential & $\sqrt{2}H_c \xi \lambda / c \tau $ & 111 $\mu$V
 \\
 Conductivity & $c^2 / 4\pi\kappa^2D$ & 31 $(\mu\Omega \text{ m})^{-1}$
 \\
 \hline
\end{tabular}\end{center}
\end{table}

\begin{table}[b!]
\caption[]{Material parameters used in the simulations. $T_\mathrm{c}$: superconducting transition temperature; $v_\mathrm{F}$: Fermi velocity; $m_\mathrm{e}$: electron mass.}
\label{tab:params_1}
\footnotesize
\begin{center}\begin{tabular}{|l|c|c| }
 \hline
 Parameter & Denotation & Value for Nb\\
 \hline
 Electron mean free path & $l$ & 6 nm
 \\
 Fermi velocity & $v_\mathrm{F}=\sqrt{2E_\mathrm{F}/m_\mathrm{e}}$ & $600$\,km/s
 \\
 Diffusion coefficient & $D=lv_\mathrm{F}/3$ & $12$\,cm$^2$/s
 \\
 Normal conductivity\,\cite{May72jap,Dob12tsf} & $\sigma=l/[3.72\times 10^{-16}\;\Omega\,\text{m}^{2}]$ & 16 $(\mu\Omega \text{ m})^{-1}$
 \\
  Relative temperature & $T/T_\mathrm{c}$ & 0.952
 \\
 Penetration depth & $\lambda=\lambda_0\sqrt{\xi_0 / (2.66l(1-T/T_\mathrm{c}))}$ & 278 nm
 \\
 Coherence length & $\xi=0.855\sqrt{\xi_0 l / (1-T/T_\mathrm{c})}$ & 60 nm
 \\
 GL parameter & $\kappa=\lambda/\xi$ & 4.7
 \\
 \hline
\end{tabular}\end{center}
\end{table}


%

\end{document}